\documentstyle[aps,graphicx,multicol]{revtex}
\begin{document}

\title{Resampling methods for document clustering}
\author{D. Volk${}^{1}$ and M.G. Stepanov${}^{1,2}$}
\address{${}^{1}$ Physics of Complex Systems, Weizmann Institute of Science,
Rehovot 76100, Israel\\
${}^{2}$ Institute of Automation and Electrometry, Novosibirsk 630090, Russia}
\maketitle
\begin{abstract}
We compare the performance of different clustering algorithms applied
to the task of unsupervised text categorization. We
consider agglomerative clustering algorithms, principal direction
divisive partitioning and
(for the first time) superparamagnetic clustering with several
distance measures. The algorithms have been applied to test
databases extracted from the Reuters-21578 text categorization test
database. We find that simple application of the different clustering
algorithms yields clustering solutions of comparable quality.
In order to achieve considerable improvements of the clustering results
it is crucial to reduce the dictionary of words considered in the
representation of the documents. Significant improvements of the
quality of the clustering can be obtained by identifying discriminative
words and filtering out indiscriminative words from the dictionary. We
present two methods, each based on a resampling scheme, for selecting
discriminative words in an unsupervised way. 
 

\end{abstract}

\begin{multicols}{2}

\section{Introduction}
Automatic text categorization has many interesting applications in
science and business. For instance formatting the
results of a web-search query, sorting news messages according to
topics or sorting incoming email with different concerns. Irrespective
of the application two principal cases are distinguished. In the first
case the categories are known and the algorithm should assign any
document to one of the known categories. Then it is useful to teach
the algorithm the considered categories and their word fields by
using a training set of labeled documents before it will be applied to
unknown documents. The algorithmic solutions to this 
problem fall into the category of \emph{supervised} learning. In the
other case the categorization of documents has 
to be done without knowing the categories nor their number. Then the
algorithm should find a reasonable partition of the document set such
that documents in the same subset of the partition are similar and 
documents of different subsets are dissimilar. This task is termed
\emph{unsupervised} text categorization and can be handled with
clustering algorithms \cite{hartigan,jain}.  

In this work we focus on the second task only.  As an illustrative
example for an application one could think of the results obtained
from web search engines. Usually the query results are 
on several subjects and only a fraction of the documents is about
what one is interested in.  It will help the user if instead of an
unsorted list the results are presented in several folders that gather
web-documents of similar content.  Further each folder could be
characterized by a list of key words.  Then one can investigate the
mass of documents that match a query in a more efficient way and the
indicated keywords may help refining the search. In the general case
of web search results we are not supplied with any training data such
that we can only use clustering algorithms in order to classify the
links. Such a combination of a web search engine and a clustering
tool has been proposed e.\,{}g.~by Boley \cite{boley1}.  

The aims of this work are threefold. First we want to compare several
methods measuring their performance on unsupervised text
categorization. Second we want to apply superparamagnetic clustering
(SPC) \cite{bla96,bla97}, a rather new method that has so far not been
considered for text categorization. And third we want to present two
methods of unsupervised feature selection and estimate the
improvements that can be achieved by their application.

\section{Clustering algorithms}
Clustering of data is usually done in three steps: representation,
calculation of similarities and application of a clustering
algorithm. As mentioned above the aim is to provide a partition of a
data set $X$ that reflects the similarities between data points. So we
should define a similarity measure $s:X\times X \rightarrow {\bf R}$
which in turn requires a numerical representation of the data.

Usually the representation is done by constructing a vector space
spanned by a set of selected features of the data. This means that one  
defines certain features and for all data assigns numbers according to
how much the features apply. Then a data point is represented by a
vector in the feature space. 

For text categorization one commonly uses the {``bag of words''}
representation. In order to do so one enlists a dictionary
$W=\left\{w_1,w_2,\ldots,w_m\right\}$ of all the words that appear at
least once in at least two of the documents. Documents are then
represented by counting the number of occurrences of each word in the
document. One thus obtains an $n\times m$-matrix $F=\left(f_{\alpha
i}\right)$ of word frequencies. $f_{\alpha i}$ is the number of
times the word $w_i$ appears in document $x_{\alpha}$ and document
$x_{\alpha}$ is represented by the feature (row-)vector $(f_{\alpha
i})_{i=1,\ldots,m}$. Typically that feature space is very
high-dimensional and the matrix is filled very sparsely, in our case
the fraction of nonzero entries is $\approx 2\%$.  

To measure the similarity of two documents one could consider for
instance the dot product of two corresponding normalized feature
vectors. Alternatively one can of course use a dissimilarity
measure. The choice of the similarity measure has to be done carefully
and influences the performance of the clustering. See
e.g. \cite{strehl} for a comparison of some similarity measures used
for text categorization.   

For the text categorization we found useful the $l_1$- and $l_2$-norms
as well as other dissimilarity measures (see 
below). Generally, in order to avoid skewness of the data due to the  
different length of the documents, it is helpful to normalize the data
such that the length of a row vector is one. 

Finally, given the similarity measure $s$ the task of a clustering
algorithm is to compute a clustering solution, i.e. a partition of
the set of data points $X$ into subsets (clusters)
$\{C_1,C_2,\ldots,C_n\}$ such that $s(x_{\alpha},x_{\beta})$ is large
when $x_{\alpha}$ and $x_{\beta}$ are in the same cluster and
$s(x_{\alpha},x_{\beta})$ is small when $x_{\alpha}$ and $x_{\beta}$
are in different clusters.   

This is a rather fuzzy description of the aim of a clustering
algorithm and we are not going to refine that point here. A good
clustering solution can be found at different resolutions, so that a
proper optimization problem can not easily be formulated. Consider for
example the biological classification of the animals. There we find
phyla that are further subdivided into classes, orders, etc. and each
level of partitioning has some justification. If we consider a
clustering of the animals cat, dog, jellyfish, mouse  
and snake, we find dog and cat in the same cluster but well separated
from the jellyfish when we a take a look from a large distance and
consider a coarse classification. A finer classification however will
separate dog and cat into different clusters. 

For many data sets this resolution is an arbitrary parameter which has   
to be determined in accordance with the desired classification task. 
We therefore do not consider a single partition of the data set as a
clustering solution but rather a hierarchy of partitions with
increasing resolution that can be represented in a tree. 

Agglomerative clustering methods \cite{jain} successively merge two
clusters until finally all data points are united. By doing so these
methods implicitly provide 
such a tree. SPC and $k$-means provide single partitions which depend
on a resolution parameter that has to be specified. Running these
algorithms with different values of the resolution parameter yields
several partitions that can be transformed into a tree. Generally the
partitions obtained at the next higher resolution are not proper
subpartitions. In order to fix this one usually considers the
intersections of high resolution clusters and low resolution clusters. 

Each triplet of a representation, a distance measure and a clustering
algorithm is considered as a clustering method. In section
\ref{results} we specify in detail the methods we apply to the test
data. All methods we compare yield a hierarchical clustering tree.

For practical purposes it is then often important to reduce the amount 
of information in the tree, and present only some selected clusters
as the essence of the clustering. Then one can apply a search algorithm
that selects the ``most meaningful'' clusters in the tree for
presentation as the clustering result.

Applying an algorithm that searches a tree for good clusters can be
considered the fourth step of clustering and can be done in various
ways. When one is using SPC one can look at the change of the
susceptibility versus the temperature and from that function one can
determine the ``best'' resolution \cite{bla96}. But the search is not
restricted to finding an optimal single resolution. For the text
categorization task we found that the \emph{natural} classes of the
documents as classified by human readers, and so specified by the
labels, are best approximated at \emph{different} levels of
resolution. Therefore we prefer other methods 
that individually judge a single cluster as good or bad and therefore
allow picking clusters from different levels of resolution. This can
be done by measuring the stability of the cluster with respect to the
resolution parameter \cite{getz} or with respect to thinning out the
dataset by considering subsamples \cite{levine}. 

However, the unsupervised identification of good clusters is a complex
issue that we do not discuss in this paper, cp.~section \ref{eval}.

\section{Feature selection: finding discriminative words}

A crucial point within the representation of the documents that bears
some potential to improve the clustering results is the
selection of words from the dictionary. Everybody will immediately
agree that the words ``and'', ``or'', ``while'' and ``with'' are
useless for document categorization. These words are not
characteristic for the content of the document and spoil the signal to
noise ratio in the representation.

Usually words like prepositions, conjunctions, etc.\ are read from a
stoplist that contains about 400 known stop-words. They are taken out
of the dictionary, i.e. out of the feature set and, as we will show in
our results, this rejection of unwanted features yields some
improvement on the results.

So far this is not new, but this procedure, however, does only a part
of the job. A more difficult problem is to get rid of those many noise
words that are not on the stoplist.  

After the application of the stoplist we remain with 5036 and 9019
words respectively to our two test databases. On the other hand,
looking at the DSR experiments that are described below, we find
improved clustering results on the basis of 350 automatically selected
words. So more than 90\% of the words that are not on the stoplist
are not needed. More than that: they make the task harder because they
add noise to the document representation.    

Finding those good words is not easy. Whether a word is a useful feature
for document classification depends on the categories that appear in
the document set and can not be told by looking at the mere word. The
word ``Italy'' for instance might be discriminative in some set of
documents where one of the categories is tourism, but it can be a pure
noise feature in another set of documents.  

A widely used method is the application of lower and upper
thresholds for the coverage of a word, i.e. the number of documents a
word appears in. This can improve the results but the quality of the
clustering depends sensitively on these thresholds and one can not
tell in general which values of these thresholds are the best. An
alternative way to select relevant features, that has been proven
useful for the analysis of microarray data, is the application of
clustering algorithms to the feature set \cite{getz}.  

In order to identify good words and throw away the bad words in an
unsupervised way we developed two strategies that are both based on
resampling.

\subsection{Word set resampling (WSR)}
Experiments with different thresholds for minimal and maximal coverage
have shown that choosing a different word subset for
clustering the same database can strongly affect the global clustering
tree structure. Such sensitivity is probably due to ``shot noise''
arising from the finite size of the dictionary (and an even smaller
number of words that appear in a single category or a single
document). One way to eliminate such noise is taking different,
e.g. randomly chosen, word subsets from the dictionary, clustering the
documents on the basis of these subsets and then averaging the result
somehow. This is the idea of the word set resampling algorithm
described in the following.  

Let us first explain the procedure that is applied for each word subset
(``probe clustering''). We cluster the documents represented only
through the words in the subset by applying an agglomerative
clustering algorithm (see below). Each agglomeration process is
continued until the stop criterion~(\ref{stopc}) holds~\cite{strug},
where $C_1$ and $C_2$ are the two largest clusters and $|C_2|<|C_1|$:
\begin{equation}
  | C_1 | + | C_2 | \ge \left(1-{1\over {\rm e}}\right) | X |.
\label{stopc}
\end{equation}
The clustering is then considered as ``good'' if the following quality 
condition is not violated.
\begin{equation}
| X | - | C_1 | - | C_2 | < | C_2 |.
\label{good}
\end{equation}
One can see that if both conditions (\ref{stopc}) and (\ref{good})
hold, then $C_1$, $C_2$ are of comparable size and contain the
majority of the documents (there is no other aim of (\ref{stopc}) and
(\ref{good})). If the probe clustering is not ``good'' then it is
rejected. For ``good'' clusterings we calculate for every word the
entropy with respect to the two biggest clusters: 
\begin{equation}
H_i = - \left( p_{i1} \log p_{i1} + p_{i2} \log p_{i2} \right) ,
\end{equation}
where
\begin{equation}
p_{ij}= { \Big| \{x_{\alpha} | f_{\alpha i}>0 \} \cap
C_j \Big| \over \Big| \{x_{\alpha} | f_{\alpha i}>0 \} \cap
\left(C_1 \cup C_2 \right) \Big| },
\quad j=1,2.
\end{equation}
Further, at that stage we check for each pair of documents if they are
in the same cluster:
\begin{equation}
M_{\alpha \beta}=\left\{\begin{array}{l} 
1, \quad \mbox{if $x_{\alpha}$, $x_{\beta}$ are in the same cluster}; \\
0, \quad \mbox{otherwise}.
\end{array}\right.
\end{equation}
We repeat this probe clustering until we have $N_{\rm R}$ ``good''
ones. The word subsets are obtained by throwing out one randomly drawn
word from each document. In order to avoid empty documents in some
cases some of the thrown out words have to be replaced. 

After doing all subsample clusterings we average $H_i$ and
$M_{\alpha\beta}$ over all $N_{\rm R}$ ``good'' probe
clusterings. Intuitively we consider $\langle H_i \rangle$ as the
quality of the word $w_i$, and $\langle M_{\alpha\beta} \rangle$ as
similarity of the documents $x_\alpha$ and $x_\beta$ ($\langle
... \rangle$ denotes average). We throw away all the words whose
average entropy exceeds the threshold:
\begin{equation}
\langle H_i \rangle > \theta \log 2,
\label{wthresh}
\end{equation}
where $\theta \in [0,1]$.

Next, we find the maximal value
$M=\max_{\alpha\ne\beta} \langle M_{\alpha\beta} \rangle$ and merge
together all the documents $x_\alpha$, $x_\beta$ with $\langle
M_{\alpha\beta}\rangle = M$.

In the next step we remove all the words that appear solely in
documents of one cluster.

Unless we are left with one huge cluster of all the documents we will
repeat this procedure. In doing so we consider as single documents in
the next step those that were merged in the current step. 

The parameters of this algorithm are $N_{\rm R}$ and $\theta$.

\subsection{Document set resampling (DSR)}

This method is motivated by the good results obtained with
WSR. It is meant to be an alternative that does not require such a
high computing time as WSR. Whereas WSR calculates the entropy of the
words 
only at a single stage of the subsample clustering, DSR tries to get
hints for the good words by looking at the whole clustering of the
subsamples. 

DSR is based on two assumptions that have been found to be true in our
experiments: first we found that the really useful words have a
considerable coverage, i.~e.~these words appear in many
documents. Second we assume that in 
agglomerative clustering, when we consider the number of mergings that
have been done as a monotonically decreasing measure of the clustering 
resolution, the cluster entropy of the good words often decreases 
earlier with decreasing resolution than the cluster entropy of the bad
words. The detailed description of the algorithm is as follows:

We create $N_{\rm R}$ subsamples $X^{(k)} \subset X, \quad
k=1,\ldots,N_{\rm R}$, each consisting of $n_{\rm sub}$ randomly
drawn documents.

To each subset $X^{(k)}$ we apply an agglomerative clustering
algorithm. We number the successive mergings of the 
agglomeration and we will refer to the intermediate clustering
solutions at step $r$. By $r=1$ we refer to the initialization with
each single document being in a separate cluster and $r=n_{\rm sub}$
corresponds to the final step where all documents are in a single
cluster.  

We then consider the words that appear in at least $n_{\rm min}$
documents of the current subsample and for each of these words we 
calculate the development of its cluster entropy in the early steps of
the agglomeration process, i.e. $r=1, 2,\ldots, 0.7n_{\rm sub}$.

Thus at step $r$ we find the clusters $C_l$ with the labels
$l \in L_r=\{1,\ldots,n_{\rm sub}-r+1\}$ and we calculate 
\begin{equation}
\tilde{H}^{(k)}_{i}(r)=-\sum_{l\in L_r} P^{(k)}(l| w_i) \log
P^{(k)}(l| w_i)  
\end{equation}
where
\begin{equation}
P^{(k)}(l| w_i)= \sum_{\alpha \in C_l} f_{\alpha i} \left/ \sum_{\alpha
\in X^{(k)}} f_{\alpha i} \right. .
\end{equation}

In order to highlight the effect of successive mergings on the cluster
entropy we normalize the cluster entropies with respect to the initial
cluster entropies at step $r=1$ 
\begin{equation}
H^{(k)}_{i}(r)=\tilde{H}^{(k)}_{i}(r)/\tilde{H}^{(k)}_{i}(1) .
\end{equation}

Now $H^{(k)}_{i}(r)$ decreases as the resolution becomes lower with
successive mergings, i.e. as $r$ increases, and finally, when all
documents have been merged to the same cluster the entropy is zero for
all words.

We found that if we consider only the words the entropy of which
decreases early in the agglomeration process we find a higher fraction
of good words that have more value for the document
clustering. However, the entropy of words that appear only in two or 
three documents naturally decreases to zero within two or three
mergings and thus (again) produces some sort of shot noise that spoils
the statistics. Thus we consider only the words that appear in at
least $n_{\rm min}$ documents of the subsample and we keep track of
the number of those words that have low entropy, i.e. we count 
\begin{equation}
q^{(k)}(r)=\left| \{i| H^{(k)}_{i}(r) < \theta\}\right|.
\end{equation}

At first the number of low-entropy-words $q^{(k)}_r$ increases slowly 
and later increases in larger steps. We consider the first increment 
\begin{equation}
\Delta^{(k)}(r) = q^{(k)}(r+1) - q^{(k)}(r)
\end{equation}
that is significantly higher than the average increment as a cutoff
criterion at which we decide to keep the words that have low entropy
according to (\ref{lec}) at that stage. That is we look for the
\emph{minimal} value $r^{(k)}_*$ fulfilling 
\begin{equation}
\Delta^{(k)}(r^{(k)}_*) > \langle\Delta^{(k)}\rangle +
\sqrt{
\left\langle (\Delta^{(k)}- \langle\Delta^{(k)}\rangle)^2\right\rangle
}
\end{equation}
and we select good words as
\begin{equation}
W^{(k)}_{\rm good}=\{w_i | H^{(k)}_{i}(r^{(k)}_*) < \theta\}.
\label{lec}
\end{equation}

Finally we consider the union 
\begin{equation}
W_{\rm good}=\bigcup_{k=1,\ldots,N_{\rm R}} W^{(k)}_{\rm good}
\end{equation}
as our selected dictionary and we cluster all documents in $X$ using
the words (features) in $W_{\rm good}$. 

The parameters of this algorithm are $N_{\rm R}$, $n_{\rm sub}$,
$n_{\rm min}$ and $\theta$.

\section{Comparison of the clustering methods}
\label{results}

\subsection{The test dataset and the different experiments}
In order to compare the performance of the different clustering
algorithms we extracted two test data sets from the known
Reuters-21578 test database for text categorization \cite{reut}. This 
database contains 21578 Reuters news messages that were manually
labeled as belonging to certain categories. We use the labels
not for the clustering procedure but in order to evaluate the quality
of the clustering results as will be described in the next section. 

For each of our two test datasets we took all documents of eight
selected categories. In order to keep things simple we considered some 
preprocessing of the labels as helpful. Those few documents that have
been labeled as belonging to more than one category were assigned
unambiguously to the first label given in the database. One should
keep in mind that the labels were given by human readers and are
therefore subject to individual perception.   

The resulting test databases, i.e. the categories that are to be
separated from one another and the number of documents extracted from
the Reuters database are listed in table~\ref{database}. Please note
that some categories appear in both tasks. This is meant as shedding
some light onto whether the good or bad separability of a category is
an individual property of that category and its word field or if it
rather a question of interference with another category using the same
words. 

We found that the quality of clustering results is sensitive to the
input dataset, particularly to the composition of the news categories
that are used. For instance, in clustering experiments done with the
first database all clustering algorithms separate documents of the
category ``coffee'' much better than they separate documents of the
category ``oilseed''. We thus concluded that the quality of the
clustering results depends on the categories and the distribution of
documents rather than on the individual documents. Obviously
clustering documents of a certain category is easier if there is a set
of discriminative words that are used in most documents of that
category and only in those documents.   

Nevertheless the quality of the clustering results varies also when
the same algorithm is applied to different randomly chosen subsets of
the database with specified numbers of documents from each category.   
In order to estimate the performance of the clustering algorithms on
the clustering task we therefore create 50 different subsets of the test
database each of which contains a specified number of documents from
each category. The clustering accuracy is then averaged among the
50 individual realizations of each experiment.

In particular we constructed four experiments for each database. We
chose 50 different subsets of 200, 500 and 800 documents preserving
the ratios of the number of documents in the categories and in another 
experiment the number of documents was the same in each category, see
table \ref{experiments}. 

\end{multicols}

\noindent\begin{center}\begin{minipage}{12cm}
\begin{table}
\begin{tabular}{|l||l|l|l|l|l|l|l|l|}
category & coffee & cpi & gnp & money-supply & oilseed & ship & sugar &
veg-oil \\
\hline
no. of docs & 124 & 75 & 117 & 113 & 78 & 204 & 145 & 93 \\ 
\hline
\hline
category & trade & crude & grain &  money-supply & interest & ship & sugar & 
money-fx \\
\hline
no. of docs & 441 & 483 & 489 & 113 & 263 & 204 & 145 & 574 \\
\end{tabular}
\caption{\label{database}The composition of the two test databases.}
\end{table}
\end{minipage}\end{center}

\noindent\begin{center}\begin{minipage}{12cm}
\begin{table}
\begin{tabular}{|l||l|l|l|l|l|l|l|l|}
experiment & coffee & cpi & gnp & money-supply & oilseed & ship & sugar &
veg-oil \\
\hline
200 & 26 & 16 & 25 & 24 & 16 & 43 & 30 & 20 \\
500 & 65 & 40 & 62 & 60 & 41 & 107 & 76 & 49 \\
800 & 105 & 63 & 99 & 95 & 66 & 172 & 122 & 78 \\
EQ & 64 & 64 & 64 & 64 & 64 & 64 & 64 & 64 \\
\hline
\hline
experiment & trade & crude & grain &  money-supply & interest & ship & sugar & 
money-fx \\
\hline
200 & 33 & 36 & 36 & 8 & 19 & 15 & 11 & 42 \\
500 & 81 & 89 & 90 & 21 & 48 & 38 & 27 & 106 \\
800 & 130 & 143 & 144 & 33 & 78 & 60 & 43 & 169 \\
EQ & 100 & 100 & 100 & 100 & 100 & 100 & 100 & 100 \\
\end{tabular}
\caption{\label{experiments}The composition of the test datasets in the
experiments.}
\end{table}
\end{minipage}\end{center}

\begin{multicols}{2}

\subsection{The clustering methods}

The representation of the documents was unchanged for all clustering
methods. We use the bag of words representation described in the first
section and applied a stoplist throwing out common words like ``and'',
``then'' or ``but''. After application of the stoplist the number of 
words is 5036 for the first database and 9019 for the second. In order
to see the improvement achieved by the application of the stoplist we
also clustered the documents using the ``noisy'' representation of the
full dictionary. These cases are indicated by the letter ``R'' in
tables \ref{1stDB} and \ref{2ndDB}. 

We applied the following clustering algorithms:

{\bf ARG} is an agglomerative method that is inspired by some ideas
from Renormalization Group theory, a similar procedure can be found in
\cite{fish}. 
It uses the dot-product of two $l_2$-normalized feature vectors as
similarity measure. The $n\times n$-matrix of pairwise similarities of
the data points is calculated in the beginning. Every single data
point is considered a cluster. The algorithm then successively joins
the two clusters which have highest pairwise similarity. Similarity of
the new cluster to another cluster is calculated from the similarities
of the joined clusters as follows:  
\begin{equation}
s(C_{\rm new},C_i)=\sqrt{0.5(s^2(C_{\rm old 1},C_i) + s^2(C_{\rm old
2},C_i))}.  
\end{equation}

{\bf AIB} is an implementation of the agglomerative information
bottleneck algorithm \cite{tishby1,tishby2}. 
Here the feature vectors are normalized with respect to the $l_1$-norm
and they are interpreted as discrete probability distribution
functions. Each entry in the feature vector gives the probability of
getting the corresponding word when one word is grabbed randomly from
the text.  
The motivation of the information bottleneck principle is
to successively join document clusters such that the loss of the
mutual information between the cluster assignment of the documents
and the occurring words is minimal.
This algorithm has been applied to unsupervised text categorization.  



{\bf SPC} is inspired by a model in theoretical physics. From
the data a Potts spin model of an inhomogenous ferromagnet is
constructed that inherits the structure of the data. The
increasing fragmentation of domains of parallel Potts 
spins when the model magnet is simulated at increasingly higher
temperatures yields a hierarchical cluster structure of the data
\cite{bla96,bla97}. We have applied this method using three
different dissimilarity measures on the data, i.e. the $l_1$- and
$l_2$-distance measures and the Jensen-Shannon-divergence
(JSD). Normalization of 
the feature vectors has been done also according to the corresponding
distance measure, i.e. $l_1$ and $l_2$ resp.~and $l_1$ for the JSD.
Of these three alternatives we obtained the best results using the
JSD. Though not very sensitively the results depend on the parameter
$k$ of the SPC algorithm which determines the number of bonds in the
model magnet \cite{bla96,bla97}. We found that the optimal value of
$k$ depends on the size of the dataset. We used $k=10$ for $n=200$,
$k=15$ for $n=500$ and $k=20$ for $n=800$ to get the best results. The
general dependence of the performance on the value of $k$ is complex 
and remains for further investigation. 

{\bf PDDP} has been proposed for text categorization by Boley
\cite{boley2}. It is a very fast method that scales linearly with the
number of documents. Here no distance measure is calculated but the
document set is recursively split into two pieces. The two subsets are
separated by a hyperplane that passes through the mean and is
perpendicular to the direction of maximal variance of the data. The
splitting of the clusters is iterated until a prescribed number of
clusters, here we chose 64, has been reached.  

\end{multicols}

\noindent\begin{center}\begin{minipage}{13cm}
\begin{table}
\begin{tabular}{|l||l|l|l|l|l|l|l|l||l|}
category & coffee & cpi & gnp & m-sup & oilseed & ship & sugar &
veg-oil & mean\\
\hline
WSR$_{64}$ 800 & {\bf 0.931} & {\bf 0.880} & 0.767 & 0.830 & 0.508 &
{\bf 0.906} & 0.730 & 0.616 & 0.771 \\ 
WSR$_{64}$ 500 & 0.926 & 0.812 & 0.770 & 0.777 & {\bf 0.531} & 0.890 &
{\bf 0.839} & {\bf 0.617} & 0.770 \\ 
WSR$_{64}$ EQ  & 0.924 & 0.828 & 0.764 & 0.784 & 0.515 & 0.885 & 0.815
& 0.599 & 0.764 \\ 
DSR$_{64}$ 200 & 0.878 & 0.829 & 0.844 & 0.889 & 0.475 & 0.769 & 0.672 
& 0.580 & 0.742 \\ 
DSR$_{32}$ 800 & 0.888 & 0.842 & {\bf 0.873} & 0.901 & 0.463 & 0.839 &
0.620 & 0.510 & 0.742 \\ 
WSR$_{64}$ 200 & 0.912 & 0.766 & 0.769 & 0.702 & 0.476 & 0.847 & 0.807
& 0.607 & 0.736 \\ 
DSR$_{32}$ 500 & 0.868 & 0.810 & 0.872 & {\bf 0.905} & 0.469 & 0.805 &
0.645 & 0.529 & 0.738 \\ 
DSR$_{32}$ EQ  & 0.866 & 0.834 & 0.838 & 0.901 & 0.472 & 0.729 & 0.667
& 0.512 & 0.728 \\ 
ARG 800 & 0.914 & 0.819 & 0.722 & 0.718 & 0.453 & 0.622 & 0.752 & 0.520
& 0.690 \\ 
AIB 800 & 0.825 & 0.829 & 0.834 & 0.857 & 0.435 & 0.730 & 0.522 & 0.491
& 0.690 \\ 
ARG 200 & 0.918 & 0.776 & 0.691 & 0.694 & 0.451 & 0.632 & 0.771 & 0.546
& 0.685 \\ 
ARG 500 & 0.916 & 0.786 & 0.709 & 0.661 & 0.468 & 0.613 & 0.769 & 0.538
& 0.682 \\ 
ARG EQ  & 0.901 & 0.817 & 0.665 & 0.697 & 0.503 & 0.606 & 0.738 & 0.528
& 0.682 \\ 
AIB 500 & 0.818 & 0.799 & 0.815 & 0.812 & 0.441 & 0.750 & 0.523 & 0.497
& 0.682 \\ 
AIB 200 & 0.791 & 0.811 & 0.793 & 0.783 & 0.443 & 0.718 & 0.562 & 0.521
& 0.678 \\ 
SPC$_{10}$ 200 & 0.864 & 0.860 & 0.794 & 0.810 & 0.423 & 0.517 & 0.608
& 0.539 & 0.677 \\ 
AIB EQ  & 0.803 & 0.827 & 0.804 & 0.852 & 0.445 & 0.656 & 0.516 & 0.494
& 0.675 \\ 
SPC$_{20}$ 800 & 0.877 & 0.847 & 0.863 & 0.802 & 0.476 & 0.453 & 0.595
& 0.486 & 0.675 \\ 
SPC$_{15}$  EQ & 0.862 & 0.835 & 0.846 & 0.841 & 0.471 & 0.502 & 0.529
& 0.492 & 0.672 \\ 
SPC$_{15}$ 500 & 0.853 & 0.847 & 0.831 & 0.789 & 0.454 & 0.473 & 0.606
& 0.477 & 0.666 \\ 
PDDP 800 & 0.878 & 0.730 & 0.601 & 0.617 & 0.407 & 0.720 & 0.758 &
0.477 & 0.649 \\ 
PDDP 200 & 0.850 & 0.757 & 0.609 & 0.638 & 0.418 & 0.659 & 0.717 &
0.488 & 0.642 \\ 
PDDP 500 & 0.873 & 0.721 & 0.594 & 0.612 & 0.390 & 0.695 & 0.736 &
0.476 & 0.637 \\ 
\hline
R AIB EQ & 0.759 & 0.789 & 0.753 & 0.746 & 0.388 & 0.653 & 0.473 & 0.476
& 0.630 \\ 
\hline
PDDP EQ & 0.782 & 0.709 & 0.587 & 0.504 & 0.415 & 0.536 & 0.632 &
0.463 & 0.579 \\ 
\hline
RAND 200 & 0.227 & 0.230 & 0.233 & 0.232 & 0.243 & 0.342 & 0.246 &
0.227 & 0.247 \\
RAND EQ & 0.205 & 0.201 & 0.200 & 0.207 & 0.200 & 0.201 & 0.197 &
0.186 & 0.200 \\
RAND 500 & 0.168 & 0.146 & 0.166 & 0.170 & 0.142 & 0.340 & 0.198 &
0.142 & 0.184 \\
RAND 800 & 0.154 & 0.113 & 0.150 & 0.127 & 0.115 & 0.335 & 0.171 &
0.110 & 0.159 \\
\end{tabular}
\caption{\label{1stDB}Maximal performance on the first database. The
best value for each category is printed in bold letters.}
\end{table}
\end{minipage}\end{center}

\begin{multicols}{2}

{\bf WSR} has been described in the previous section. For the ``probe
clustering'' we applied the ARG algorithm as described above. Further
we chose $\theta=0.8$, the optimal number of subsamples $N_{\rm R}$
has been determined in a supervised way (see figure 1), the
performance saturates at $N_{\rm R}=64$. 

{\bf DSR} has been used as described above. For clustering
the subsamples as well as for the final round of clustering the whole
dataset with the reduced word set we employed the AIB algorithm. We
found $n_{\rm sub}=100$ and $n_{\rm min}=5$ suitable for all
experiments. Also we used $\theta=0.8$ and again the
number of subsamples $N_{\rm R}$ has been determined
experimentally. We found good performance at about 32 subsamples (see
figure 1). However, further increasing the number 
of subsamples spoils the selection of ``good'' words and leads
to a decrease of the performance. In the limit $N_{\rm R} \rightarrow
\infty$ the word selection then degenerates to a cutoff criterion
prescribing a minimal coverage.
 
The number of ``good'' words that have been selected by the DSR method,
when being applied to the 800 documents experiment with $N_{\rm R}=32$,
is 328$\pm$29 for the first database and 365$\pm$33 for the second
database.      

{\bf RAND} has been included to provide a baseline for the evaluation
scheme. We produced trees by randomly agglomerating document clusters.

\end{multicols}

\noindent\begin{center}\begin{minipage}{13cm}
\begin{table}
\begin{tabular}{|l||l|l|l|l|l|l|l|l||l|}
category & trade & crude & grain &  m-sup & interest & ship & sugar & 
m-fx & mean\\
\hline
WSR$_{64}$ 800 & 0.645 & 0.861 & {\bf 0.767} & 0.578 & 0.468 & 0.633 &
0.796 & 0.666 & 0.677 \\ 
WSR$_{64}$ 500 & 0.627 & {\bf 0.864} & 0.750 & 0.590 & 0.499 & 0.589 &
0.804 & {\bf 0.681} & 0.676 \\ 
WSR$_{64}$ EQ  & 0.627 & 0.845 & 0.701 & 0.664 & 0.513 & {\bf 0.698} &
{\bf 0.889} & 0.453 & 0.674 \\ 
WSR$_{64}$ 200 & 0.664 & 0.861 & 0.742 & 0.662 & 0.514 & 0.541 & 0.719
& 0.652 & 0.669 \\ 
DSR$_{32}$ EQ  & 0.665 & 0.829 & 0.606 & 0.706 & 0.593 & 0.676 & 0.708
& 0.527 & 0.664 \\ 
DSR$_{32}$ 500 & 0.676 & 0.847 & 0.725 & 0.633 & 0.554 & 0.546 & 0.600
& 0.623 & 0.650 \\ 
DSR$_{32}$ 800 & 0.680 & 0.850 & 0.731 & 0.618 & 0.553 & 0.583 & 0.543
& 0.639 & 0.650 \\ 
DSR$_{32}$ 200 & 0.606 & 0.818 & 0.701 & 0.668 & 0.569 & 0.523 & 0.642
& 0.607 & 0.642 \\ 
SPC$_{20}$ EQ  & {\bf 0.697} & 0.802 & 0.473 & {\bf 0.771} & {\bf
0.656} & 0.445 & 0.629 & 0.478 & 0.619 \\ 
AIB EQ          & 0.643 & 0.748 & 0.504 & 0.704 & 0.582 & 0.632 & 0.624
& 0.471 & 0.614 \\ 
AIB 800         & 0.668 & 0.778 & 0.673 & 0.624 & 0.525 & 0.562 & 0.453
& 0.578 & 0.608 \\ 
AIB 500         & 0.661 & 0.770 & 0.668 & 0.642 & 0.548 & 0.521 & 0.472
& 0.570 & 0.606 \\ 
ARG 200         & 0.637 & 0.769 & 0.576 & 0.591 & 0.602 & 0.482 & 0.632
& 0.538 & 0.603 \\ 
AIB 200         & 0.620 & 0.734 & 0.639 & 0.647 & 0.558 & 0.501 & 0.538
& 0.568 & 0.600 \\ 
ARG 500         & 0.631 & 0.778 & 0.596 & 0.520 & 0.566 & 0.466 & 0.652
& 0.572 & 0.597 \\ 
ARG EQ          & 0.609 & 0.752 & 0.485 & 0.610 & 0.611 & 0.552 & 0.704
& 0.436 & 0.595 \\ 
ARG 800         & 0.631 & 0.774 & 0.573 & 0.494 & 0.584 & 0.478 & 0.631
& 0.576 & 0.593 \\ 
SPC$_{20}$ 800 & 0.638 & 0.699 & 0.667 & 0.588 & 0.639 & 0.442 & 0.342
& 0.540 & 0.570 \\ 
SPC$_{15}$ 500 & 0.668 & 0.730 & 0.669 & 0.576 & 0.604 & 0.435 & 0.423
& 0.559 & 0.583 \\ 
SPC$_{10}$ 200 & 0.644 & 0.671 & 0.647 & 0.611 & 0.569 & 0.455 & 0.516
& 0.548 & 0.583 \\ 
\hline
R AIB EQ & 0.632 & 0.673 & 0.418 & 0.705 & 0.544 & 0.596 & 0.591 & 0.442
& 0.575 \\ 
\hline
PDDP EQ & 0.642 & 0.522 & 0.403 & 0.571 & 0.557 & 0.473 & 0.659 &
0.409 & 0.530 \\
PDDP 200 & 0.644 & 0.587 & 0.470 & 0.521 & 0.536 & 0.424 & 0.502 &
0.452 & 0.517 \\ 
PDDP 500 & 0.689 & 0.568 & 0.471 & 0.456 & 0.551 & 0.363 & 0.456 &
0.431 & 0.498 \\ 
PDDP 800 & 0.668 & 0.543 & 0.457 & 0.419 & 0.549 & 0.345 & 0.431 &
0.396 & 0.476 \\ 
\hline
RAND 200 & 0.253 & 0.285 & 0.284 & 0.271 & 0.227 & 0.225 & 0.242 &
0.331 & 0.265 \\
RAND 500 & 0.210 & 0.245 & 0.243 & 0.158 & 0.138 & 0.138 & 0.152 &
0.332 & 0.202 \\
RAND EQ  & 0.198 & 0.191 & 0.193 & 0.189 & 0.178 & 0.187 & 0.182 &
0.174 & 0.187 \\
RAND 800 & 0.180 & 0.226 & 0.227 & 0.118 & 0.103 & 0.099 & 0.106 &
0.329 & 0.173 \\
\end{tabular}
\caption{\label{2ndDB}Maximal performance on the second database. The
best value for each category is printed in bold letters.}
\end{table}
\end{minipage}\end{center}

\begin{multicols}{2}

\subsection{Evaluation of clustering results}
\label{eval}
We now want to check if the obtained clustering solutions provide good 
estimations of the categories as given by the labels. Ideally one
wishes that a cluster contains all the documents of one category and
only these. In reality we find that a cluster has documents with
different category labels. In order to measure how close a cluster
comes to this ideal case we define two numbers which are calculated
for each cluster. The category to which the labels assign the most
documents of a cluster is called the \emph{type} of that cluster. As
\emph{purity} we define the fraction of documents of that type within
the cluster and as \emph{efficiency} we define the number of documents
the label of which is the cluster type divided by the overall number
of documents with that label.      

Let ${\cal C}_1,{\cal C}_2,\ldots ,{\cal C}_l \subset X $ be the ideal
clusters with respect to the labels, i.e. ${\cal C}_1$ contains all
the documents that are labeled as belonging to category one, and let
further be $C$ a cluster found by
the algorithm. Then the purity of $C$ is defined as $P(C)=\max_i | C
\cap {\cal C}_i | / | C |$ and the type of the cluster $T(C)$ is the
index $i$ for which the expression on the right side is maximal. The
efficiency accounts for the fraction of all documents of the category
which are gathered in the cluster: $E(C)=| C \cap {\cal C}_{T(C)} 
| / | {\cal C}_{T(C)} |$.  

In order to have one quality measure that combines these two issues we
use the commonly used $F_1$ measure which was introduced by van
Rijsbergen \cite{vrijs} and is defined as 

\begin{equation}
F_1={2PE \over P+E}.
\end{equation}

The $F_1$ measure considers purity and efficiency to be equally
important for the quality of a cluster. It is evaluated for each
cluster in the entire tree and the best clusters with respect to each
category are taken as the quality vector for that particular
clustering experiment. The quality of each tree is the mean of the
best $F_1$-values for each category. Tables \ref{1stDB} and
\ref{2ndDB} show the values of the quality vector for the different
methods and experiments. Each line in the two tables has been averaged 
among 50 different realizations of the corresponding experiment,
i.e. by application to 50 different document sets with the same
category distribution.   

\begin{figure}		
\begin{center}
\includegraphics{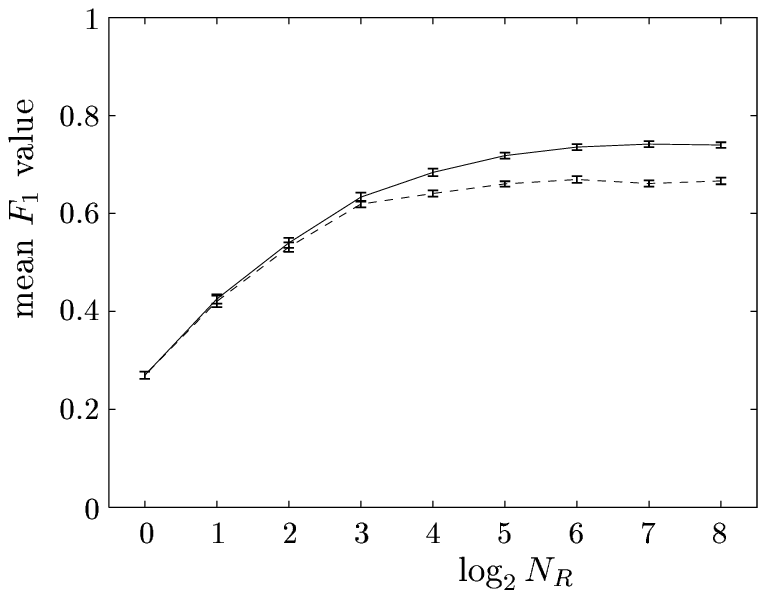}
\includegraphics{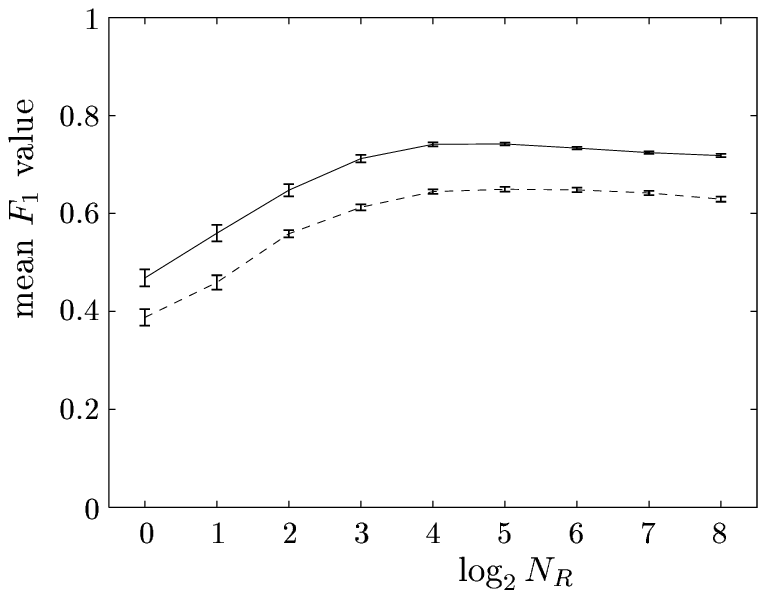}
\end{center}
\caption{Dependence of the clustering performance of the resampling
methods WSR (applied to the 200 experiment, upper graph) and DSR
(applied to the 800 experiment, lower graph) on the number of
subsamples. Saturation occurs at 64 subsamples for the WSR method. DSR
works best at $N_{\rm R}=32$. The solid line corresponds to the first
database, the dashed line to the second.}  
\label{sat}
\end{figure}

\begin{figure}
\begin{center}		
\includegraphics{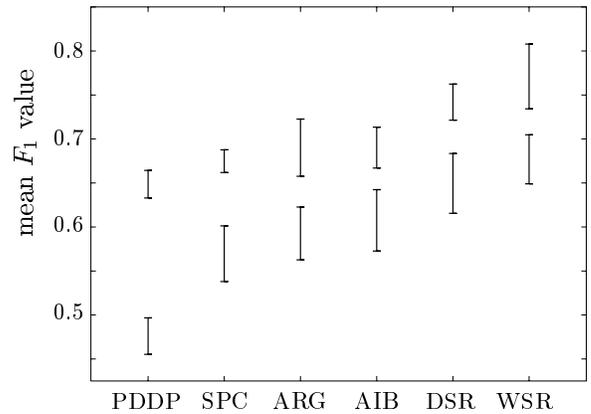}
\end{center}
\caption{Direct comparison of the clustering methods. The bars
indicate mean and standard deviation of the mean $F_1$-values obtained
by applying the different methods to the 800 documents experiments in
the first (upper bars) and second (lower bars) database.} 
\label{vergl}
\end{figure}

As mentioned above we here put aside the interesting question of how
to find the good clusters within the tree, i.e. estimate in an
unsupervised way which clusters at which resolution are good
approximations of the underlying categories. 

Thus the values presented in tables \ref{1stDB} and \ref{2ndDB} are
upper bounds for what is achievable with any such search algorithm. We
think that this problem can be separated from the basic clustering
problem as the occurrence of good clusters in the tree is of course
the prerequisite that limits all achievements of any algorithm that
searches the tree for good clusters.

\noindent\begin{center}\begin{minipage}{8cm}
\begin{table}
\noindent\begin{center}\begin{minipage}{2.8cm}
\begin{tabular}{|l|l|}
ARG & $96\pm9$ \\
DSR & $136\pm32$ \\
AIB & $286\pm54$ \\
SPC & $801\pm50$ \\
WSR & $9678\pm1192$ \\
\end{tabular}
\end{minipage}\end{center}
\caption{\label{coco}Computational cost in seconds. All values
correspond to applying the algorithms to the 800 experiment of the
first database. The CPU has been PIII with 650 MHz. PDDP was run under
MATLAB and thus cannot be compared.}  
\end{table}
\end{minipage}\end{center}



\section{Conclusions}

We find that the quality of the clustering results depends to a large
extent on the dataset. In particular we observe that the performance
as measured in this paper is almost always better on larger categories.
When comparing the results of single categories in the 800 and EQ
experiments, we find that they are better in the case where the number
of documents of that category is larger. For example in the second
database categories ``trade'', ``crude'', ``grain'', ``money-fx'' are
larger in the 800 experiment and the results are also better in the
800 experiment. All other categories are larger in the EQ experiment
and also there are the better results for these categories. Also in
this way the categories ``ship'' and ``money-supply'' can be better 
resolved in the context of the first database.

Also we think that the resolvability of a category is influenced by
interference with other categories in the dataset through an overlap
of the characteristic word fields. We believe that if the
characteristic words of a category are also used in documents of other 
categories that category can not be resolved as good as if there were
no close categories. 

Comparing the results for the  ``money-supply'', ``ship'' and
``sugar'' categories in the EQ experiments of the two databases gives
a clue to possible interference. We find that ``money-supply'' and
``ship'' are better resolved in the EQ experiments on the first
database, whereas ``sugar'' is better in the second database.

Further we observe that some categories appear to have a preferred
algorithm or vice versa. So SPC performance in the EQ experiment of
the second database peaks in categories ``trade'', ``money-supply''
and ``interest'' whereas the results for the other categories are only 
moderate.   

However, the ranking of the performance of different clustering
methods does not sensitively depend on the data. We found that the
level of performance of SPC, ARG and AIB is almost the same. Results
obtained with PDDP are not as good, whereas the advantage of this
method is that it is much faster on large databases. PDDP does not
require the calculation of a (dis-)similarity matrix and its time
consumption scales linear with the number of documents.

The feature selecting methods that we propose in this paper can
improve the results. WSR yields the highest performance but has on the
other hand a very high computational cost. As it is implemented, the
required time scales with $n^3$. DSR gives moderate improvement of the
clustering quality but is in comparison to WSR much faster. The time
consumption of DSR is dominated by the size of the subsets. Thus for
large datasets, if one can probe the discriminative words with
comparably small subsets it will be faster than SPC, AIB and ARG that
all rely on the computation of a complete distance matrix on the basis
of the whole word set. Another little advantage of the feature
selecting methods is that the application of a stoplist becomes
obsolete, WSR and DSR perform as good on the raw data matrix. 

\acknowledgments

We are grateful to E.~Domany, G.~Getz, S.~Gnutzmann and A.~Shehter for
useful discussions. This work was partially supported by the Minerva
foundation.

\end{multicols}

\end{document}